\documentclass[superscriptaddress,final,preprint,showkeys]{revtex4-1}
\usepackage{graphics}
\usepackage[dvips]{graphicx}
\usepackage{times}
\usepackage{multirow}
\usepackage[%
  colorlinks=true,
  urlcolor=blue,
  linkcolor=blue,
  citecolor=blue
]{hyperref}

\usepackage{color}

\begin{document}

\title{Nanowars can cause epidemic resurgence and fail to promote cooperation\\
\small{Hypothetical nanotech-based wars targeting non-cooperative infectious agents\\ would reduce, not improve planetary health}}

\author{Dirk Helbing}
\thanks{Electronic address: \href{mailto:dhelbing@ethz.ch}{\textcolor{blue}{dhelbing@ethz.ch}}}
\affiliation{Computational Social Science, ETH Z{\"u}rich, Stampfenbachstra{\ss}e 48, 8092 Z{\"u}rich, Switzerland}
\affiliation{Complexity Science Hub Vienna, Josefst{\"a}dterstra{\ss}e 39, 1080 Vienna, Austria}

\author{Matja{\v z} Perc}
\thanks{Electronic address: \href{mailto:matjaz.perc@gmail.com}{\textcolor{blue}{matjaz.perc@gmail.com}}}
\affiliation{Complexity Science Hub Vienna, Josefst{\"a}dterstra{\ss}e 39, 1080 Vienna, Austria}
\affiliation{Faculty of Natural Sciences and Mathematics, University of Maribor, Koro{\v s}ka cesta 160, 2000 Maribor, Slovenia}
\affiliation{Department of Medical Research, China Medical University Hospital, China Medical University, Taichung 404332, Taiwan}
\affiliation{Alma Mater Europaea, Slovenska ulica 17, 2000 Maribor, Slovenia}

\begin{abstract}
\textbf{In a non-sustainable, ``over-populated'' world, what might the use of nanotechnology-based targeted, autonomous weapons mean for the future of humanity? In order to gain some insights, we make a simplified game-theoretical thought experiment. We consider a population where agents play the public goods game, and where in parallel an epidemic unfolds. Agents that are infected defectors are killed with a certain probability and replaced by susceptible cooperators. We show that such ``nanowars'', even if aiming to promote good behavior and planetary health, fail not only to promote cooperation, but they also significantly increase the probability of repetitive epidemic waves. In fact, newborn cooperators turn out to be easy targets for defectors in their neighborhood. Therefore, counterintuitively, the discussed intervention may even have the opposite effect as desired, promoting defection. We also find a critical threshold for the death rate of infected defectors, beyond which resurgent epidemic waves become a certainty. In conclusion, we urgently call for international regulation of nanotechnology and autonomous weapons.}
\end{abstract}


\maketitle

\linespread{1.1}

\section*{\normalsize{Introduction}}
\label{intro}

In November 2019, newspapers reported ``Earth Needs Fewer People to Beat the Climate Crisis'', a call for action with more than 11.000 signatories~\cite{call}. At the same time, thousands of experts in the AI and robotics communities have raised concerns about autonomous weapons such as drones and ``killer robots''~\cite{future}. In perspective, when combined with face recognition or other identification techniques, it has been warned that ``slaughterbots'' could be used for targeted killing, e.g., of dissidents or unwanted minorities. Given that this opens new possibilities for genocide, the recent failure of the United Nations to ban autonomous weapons is concerning~\cite{converse}.

Here, we discuss another conceivable scenario that has not been talked about much: In a simplified though experiment, we will study some implications of a scenario assuming that nanotechnology might be used for targeted killing and population control outside of classical battlefields. For this purpose, we apply game-theoretical methodology, which has been used to analyze situations of competition and war since the very beginning, and extend it by an epidemiological spreading model.

Without any doubt, we are now living in digital times, where huge amounts of Big Data are created by the Internet of Things (IoT), based on sensor networks, from which data can be obtained in wireless ways. By now, the technology has advanced to the point, where nanotechnology can be used to establish what is called the ``Internet of Bodies'' (IoB)~\cite{IoB}. In fact, the World Economic Forum (WEF) stated in June 2020~\cite{wef}: ``We're entering the era of the `Internet of Bodies': collecting our physical data via a range of devices that can be implanted, swallowed or worn.''

Applications like the IoB may be based on the exposition of bodies to nanoparticles, i.e. less than micrometer-scale particles or structures, not visible to the naked eye, which may be used, for example~\cite{rand, INT},
\begin{enumerate}
\item to give people a unique electronic identity (e-ID) and track them,
\item to measure body functions and cellular activities, or
\item to manipulate processes in the body (e.g., to heal a disease).
\end{enumerate}

Of course, the opposite, i.e. causing a disease would also be possible. Hence, people could be killed slowly but surely. This is particularly evident for deadly viruses, which are particular kinds of nanoparticles and might be modified by gain-of-function research~\cite{functionalgain}. For example, it is conceivable to wirelessly switch certain functionality on or off~\cite{onoff}. For further information one may search for the ``Internet of (Bio-)Nano Things'', ``of Medical Things'', ``of Military Things'', and ``of Virus Things''.

Inspired by this, we discuss the following simplified scenario that does not claim realism, but rather serves to initiate an overdue debate. In the very tradition of ``thought experiments'' such as ``Schr{\"o}dinger's cat''~\cite{thoughtexperiment}, the scenario discussed here is purely hypothetical and does not claim similarities with real events. It innovatively combines the spatial disease spreading of viruses according to a SIRS model~\cite{sirs2} with the dynamics of cooperation or defection in a social dilemma game~\cite{nowak_s06}. We further assume that the hypothetical kinds of viruses can be switched between two different states: a harmless, mild, symptom-free one and an aggressive, potentially deadly one.

In other words, the assumed kinds of viruses may be weaponized in a targeted and personalized way. Inspired by the ``Karma Police'' program~\cite{intercept}, which judges the value of a person with a citizen score that is determined by mass surveillance, we further assume that switching could be targeted, using personal data, as in the ``Skynet'' program~\cite{skynet}.

Our model corresponds to a setting in which a simplified kind of citizen score is determined by the behavior of a person, namely where a positive value is given to a cooperative agent, while a defector gets a negative value. Moreover, it is assumed that a negative value may trigger a triage decision, which turns the virus deadly when the targeted agent is infected.

We assume that the targeting, i.e. the personalized switching between the harmless and deadly state of the hypothetical, ``customized'' virus, is controlled by an autonomous AI system, which has been tasked to ``maximize planetary health''. In our model, the system targets and kills infected defectors with the intention to ``create a paradise full of healthy cooperators''. We will show that such a scenario fails to reach its goal of eliminating defectors who show antisocial behavior, and it further tends to cause epidemic resurgence, i.e. repetitive epidemic waves.

\section*{\normalsize{Mathematical model}}
\label{model}

As basis of our model, we use a square lattice of size $L^2$ with periodic boundary conditions, where each agent is connected to its four nearest neighbors. The public goods game is thus played in overlapping groups of size $G=5$, where each agents belongs to $g=G$ different groups~\cite{santos_n08}. Cooperators and defectors are initially distributed uniformly at random, and subsequently cooperators contribute a fixed amount, here considered to be equal to $1$ without loss of generality, in each instance of the public goods game, while defectors contribute nothing. The sum of all contributions in each group is multiplied by the factor $R>1$, and the resulting public goods are distributed amongst all group members irrespective of their initial contribution. If $s_x = C$, i.e. player $x$ is cooperative, the payoff of $x$ from every group $g$ is
\begin{equation}
P_{C}^g= R N_{C}^g/G - 1 \, ,
\end{equation}
where $N_{C}^g$ is the number of cooperators in group $g$.
If $s_x = D$, i.e. player $x$ is defecting, the payoff of $x$ is
\begin{equation}
P_{D}^g= R N_{C}^g/G \, .
\end{equation}

Monte Carlo simulations~\cite{binder_88} of this public goods game are carried out based on the following elementary steps: A randomly selected player $x$ plays the public goods game as a member of all the $g=1, \ldots, G$ groups of size $G$, whereby its payoff is
\begin{equation}
P_{s_x} = \sum_g P_{s_x}^g \, .
\end{equation}
Next, player $x$ chooses one of its nearest neighbors at random, and this player also acquires its payoff $P_{s_y}$ in the same way. Finally, player $y$ copies the strategy $s_x$ of player $x$ with a probability given by a multinomial logit model. This is represented here by the Fermi function
\begin{equation}
w(s_x \to s_y)=1/\{1+\exp[(P_{s_y}-P_{s_x})/K]\} \, ,
\end{equation}
where $K$ quantifies the uncertainty in strategy adoptions due to errors in decision making or incomplete information. We use $K=0.5$ without loss of generality, and in agreement with the most commonly used setup~\cite{perc_pr17}. Each full Monte Carlo Step (MCS) entails $L^2$ such elementary steps, as $L^2$ is the size of the periodic lattice, which gives a chance to every player to change its strategy once, on average, in one full MCS.

\begin{figure*}[t]
\centering{\includegraphics[width = 16cm]{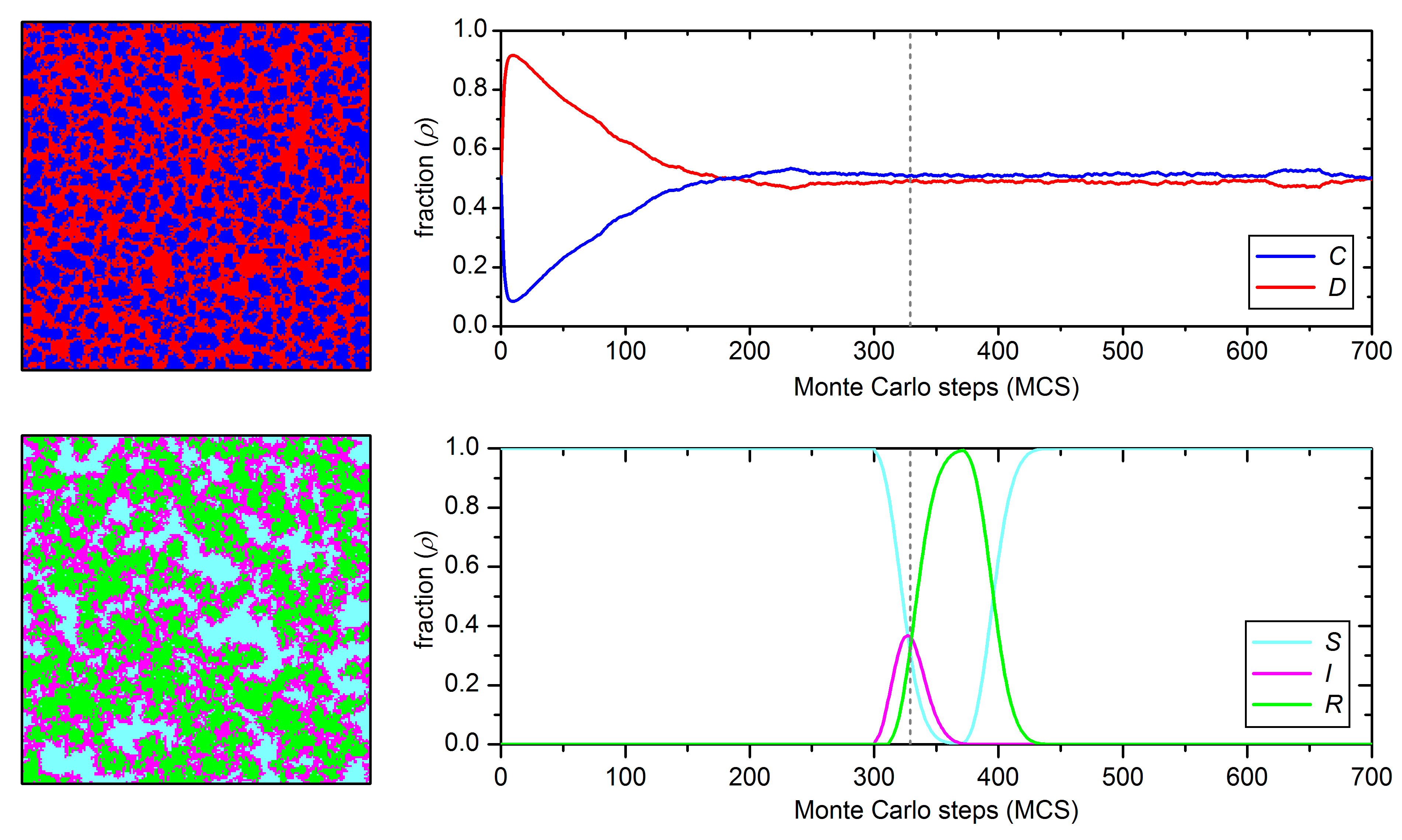}}
\caption{\textbf{Illustration of the model components of the nanowar thought experiment.} \textbf{Upper row:} Spatiotemporal dynamics of the public goods game, as obtained for the multiplication factor $R=4$. The snapshot shows the spatial distribution of cooperators (blue) and defectors (red), as obtained at $329$ MCS (dashed vertical line in the right plot). It can be observed that cooperators form compact clusters in the ``sea of defectors'', which is why they can survive even below the survival limit for well-mixed populations, $R=G$. The time course of the fractions of the two strategies shows that $\rho_{C}$ hovers slightly above $50\%$ in the stationary state, which is reached after approximately $250$ MCS. \textbf{Lower row:} Spatiotemporal dynamics of the SIRS model that runs concurrently with the public goods game, as obtained for the parameters described in the section on the Mathematical Model. %Section~\ref{model}.
The snapshot shows the spatial distribution of susceptible (cyan), infectious (magenta), and recovered (green) agents, as obtained at the same point in time as the public goods game snapshot (dashed vertical line in the right plot). The compact fronts of infectious individuals spreading into susceptible regions can be clearly observed, as well as the recovered agents left in their wake. The time course of the fractions of the three states shows a typical epidemic wave starting at $300$ MCS, after $1\%$ of the population was infected uniformly at random. The peak in the fraction of infectious individuals follows the peak in the fraction of recovered individuals, which reaches $100\%$ of the population, indicating that practically all agents have been infected once during the wave. Finally, the population reaches an all-susceptible state again after the last of the recovered lose their immunity after $t_i$ time steps.}
\label{making}
\end{figure*}

In what follows, we will use between $L=50$ and $400$ and three different values of the multiplication factor $R$ that describe three significantly different outcomes of the public goods dilemma. For $R=3$ we have a complete absence of cooperators, such that their average density in the stationary state, when the average fraction of both strategies becomes time independent, is $\rho_{C}=0$. For $R=4$ we have a mixed cooperation-defection state of the stylized society, such that the average fraction of cooperators is $\rho_{C}=0.52$ (see upper row of Fig.~\ref{making} for details). And finally, for $R=5$ we have a near domination of cooperators, such that their average fraction is $\rho_{C}=0.93$. These are well-known results, which have been reported often before~\cite{perc_pr17}.

Simultaneously to the public goods game, the same $L^2$ agents on the square lattice are assumed to be subject to an epidemic, which we simulate by means of Monte Carlo simulations of the susceptible-infectious-recovered-susceptible (SIRS) model~\cite{farkas2002mexican, moreno2002epidemic, pastor_rmp15, wang_z_pr16}. Accordingly, each agent can be either in the susceptible ($e_x=S$), infectious ($e_x=I$), or recovered ($e_x=R$) state. Initially all agents are placed in the $S$ state, but when the public goods game reaches the stationary state, we assume that $1\%$ agents have been randomly infected according to a uniform distribution, which makes them infectious, thus setting off an epidemic wave.

The following elementary steps apply: We first select an agent $x$ uniformly at random. If $e_x=S$, we then select one of its neighbors $y$ uniformly at random and, if $e_y=I$, player $x$ becomes infected with probability $p$. We here use $p=0.5$ without loss of generality~\cite{pastor_rmp15, wang_z_pr16}. If the neighbor $y$ is in any other state, nothing happens. On the other hand, if $e_x=I$, we verify whether at least $t_r$ full MCS have passed since it became infected. If yes, agent $x$ becomes recovered, and if not, it remains infected. We note that $t_r$ is the so-called recovery time, which we set to $10$ full MCS, such that combined with $p=0.5$, roughly each agent on the square lattice is infected once during the first epidemic wave (see the green line in the bottom row of Fig.~\ref{making}). Lastly, if $e_x=R$, we verify whether at least $t_i$ full MCS have passed since it became recovered. If yes, agent $x$ becomes susceptible again, and if not, it remains recovered. Similarly to $t_r$, $t_i$ is the immunity time, i.e., the time after recovery during which an agent cannot become reinfected, which we set to $6 t_r$. This implies that, in the absence of any further triggered infections, the first wave of the epidemics is almost surely also the last one (see lower row of Fig.~\ref{making} for details), as all infectious agents become recovered before those who were infected first become susceptible again.

To introduce and study possible impacts of a targeted nanowar, we finally assume that it is possible to turn the virus deadly, such that, with probability $q$, a infectious defector dies after $3/4$th of the recovery time. To give this intervention the best chance to have a ``prosocial'' effect, we assume that deceased infectious defectors will be replaced by newborn susceptible cooperators. Hence, in resemblance to the Biblical ``Last Judgment'' scenario, some entity (e.g. AI system) is trying to decrease the number of infectious agents and to rid the population of freeriders.

\section*{\normalsize{Results}}
\label{results}

\begin{figure*}[t]
\centering{\includegraphics[width = 16cm]{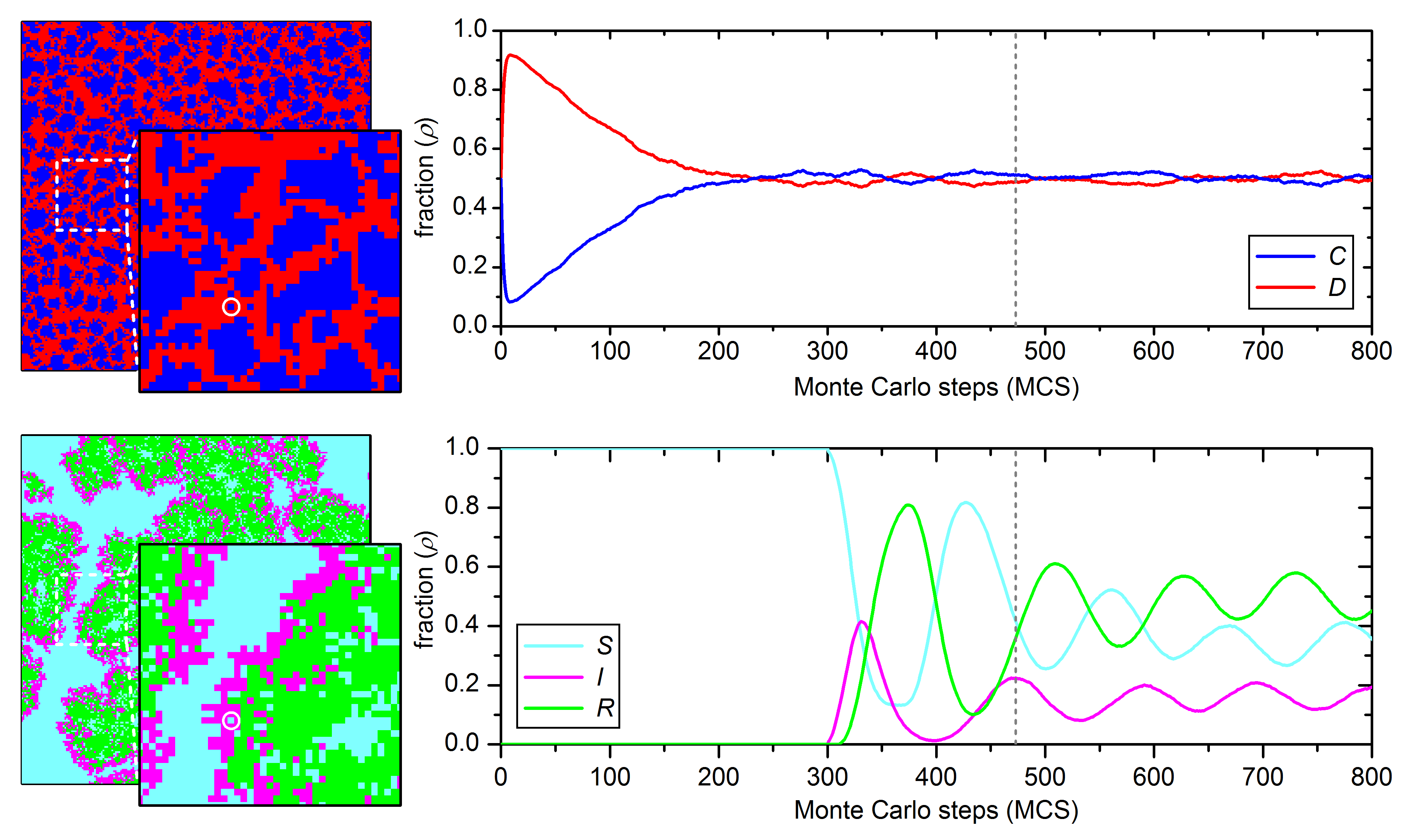}}
\caption{\textbf{Nanowar unleashed.} We show the same results as in Fig.~\ref{making}, just with the difference that here an autonomous AI system is assumed to target defectors, i.e. the ``nanowar machine'' is turned on. Our simulations use $q=0.5$, meaning that there is a $50\%$ chance that an agent who is an infected defector will be killed $3/4$th into its recovery time $t_r=10$ MCS. Subsequently, all emerging vacant sites are filled with susceptible cooperators. While one might expect that this creates a world full of healthy cooperators over time, the upper row shows that the public goods dynamics is only marginally affected, i.e. defection does not disappear. Rather counterintuitively, given that defectors are eliminated and replaced with cooperators, the average stationary fraction of cooperators is even somewhat reduced. It turns out that the newly introduced cooperators (see the white circle in the enlarged snapshot), which are often isolated, are easy targets for defectors, which can therefore spread somewhat more effectively than in the absence of nanowar. The lower row, on the other hand, reveals further trouble caused by nanowar, namely repetitive waves of infection that never seize. The reason for this unfavorable development are the newly placed susceptible cooperators (see the white circle in the lower enlarged snapshot), which keep the epidemics alive and do not allow the population to develop full-scale immunity.}
\label{unleash}
\end{figure*}

In what follows, we explore the impact of different values of $q$---akin to the intensity of nanowar---on public cooperation and epidemic spreading for the three different values of $R$. In Fig.~\ref{unleash}, we first revisit the results presented in Fig.~\ref{making} for $q=0.5$, thus unleashing nanowar with a $50\%$ chance of survival, while an infected defector will be killed and replaced by a susceptible cooperator with a probability of 50\%. Fundamental differences, especially in the SIRS dynamics (bottom row), can be observed immediately. Instead of a single epidemic wave, we have concurrent waves that never seize. In this way, the nanowar has a self-perpetuating nature: newly emerging susceptible cooperators feed new epidemic waves, which eliminate reemerging defectors.

The public goods dynamics is comparatively less affected, although it is clear that the nanowar entirely misses its assumed goal of promoting planetary health and social welfare through increased cooperation. On the contrary, if anything, the level of cooperation even somewhat decreases, because the newly introduced cooperators enable surrounding defectors to exploit them. In addition to the never-ending epidemic waves, we thus observe a decline of public cooperation due to the counterintuitive effect that newborn cooperators do not increase cooperation, but rather promote nearby freeriding.

\begin{figure*}[t]
\centering{\includegraphics[width = 16cm]{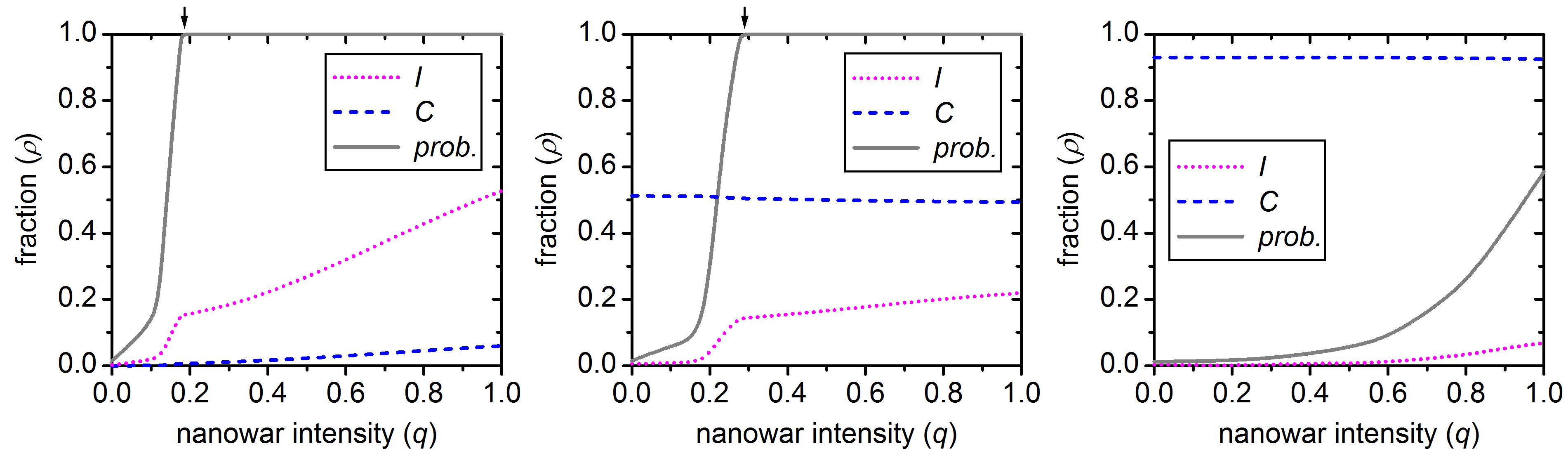}}
\caption{\textbf{Quantitative consequences of nanowar.} From left to right we show the average stationary fraction $\rho_C$ of cooperators, the average stationary fraction $\rho_I$ of infectious agents, as well as the probability for the emergence of recurrent epidemic waves (prob.) in dependence on the nanowar intensity $q$, as obtained for $R=3$, $4$, and $5$. For $R=3$---corresponding to all out defection---we have a critical value of nanowar intensity equal to $q_c=0.175(4)$ (marked by the short black arrow), beyond which endless epidemic waves are certain. As $q$ increases, we also observe a rise in $\rho_I$ as well as a slight increase in $\rho_C$. For $R=4$, corresponding to a mixed $C+D$ state, we have $q_c=0.276(4)$ as well as an increased level $\rho_I$ of infectious agents, and some decrease in the level $\rho_C$ of cooperators as we increase the value of $q$. Lastly, for $R=5$, corresponding to an almost all out cooperation, the probability for endless epidemic waves always stays below $1$, reaching a maximum of just below $0.6$ when $q=1$. Nonetheless, as nanowar intensity $q$ is increased from $0$ to $1$, one can see that the level $\rho_I$ of infectious agents increases, while the level $\rho_C$ of cooperators somewhat decreases.}
\label{quanty}
\end{figure*}

It is of further interest to determine more accurately to what degree these worrying consequences of nanowar manifest for other values of $q$ (the nanowar intensity), and for different propensities of the environment to support cooperation. We therefore vary $q$ with $0 \leq q \leq 1$ and record the average stationary fraction $\rho_C$ of cooperators, the average stationary fraction $\rho_I$ of infectious agents, and the probability for the emergence of recurrent epidemic waves over $300$ independent realizations of the nanowar scenario. The results shown in Fig.~\ref{quanty} essentially confirm the worst case, namely epidemic resurgence and failure to promote cooperation---with the exception of a slightly positive trend of cooperation for $R=3$ (leftmost panel). It should be noted, however, that for $R=3$ the default state in the public goods game is all out defection. Accordingly, the little rise in $\rho_C$ as $q$ increases is simply due to the fact that, when defectors are killed, they are replaced by cooperators, which results in an artificially maintained, low, non-zero fraction of cooperators. The outcome, however, is not the emergence of compact cooperative clusters, but short-lived isolated cooperators that turn into defectors, only to be killed and replaced again by the nanowar machinery. The cost for the Pyrrhic victory in this brave new world is half of the population always being in an infectious state.

Results in Fig.~\ref{quanty} also show that, for larger $R$, i.e., in more cooperative environments, the negative impacts of nanowar are likely to be smaller---in that recurrent epidemic waves are lower and flatter, as well as never fully certain, which ultimately results in a lower value $\rho_I$ of infectious agents. For example, for $R=4$, a full-on nanowar with $q=1$ leaves only half as much of the population ($22\%$) in the infectious state compared to $R=3$ ($53\%$). For $R=5$, this further decreases to just $6.9\%$. This is understandable, of course, since the lack of defectors in predominantly cooperative societies drastically decreases the number of available targets for the nanowar machine. Therefore, the negative consequences that we have reported for non-cooperative ($R=3$) and somewhat cooperative ($R=4$) societies, do not manifest so strongly, but are still there.

Furthermore, with regard to the results presented in Fig.~\ref{quanty}, we note that the critical threshold in $q$ for never ending epidemic waves also increases with increasing $R$. For $R=3$, a nanowar with intensity $q_c=0.175(4)$ is already guaranteed to always induce recurrent epidemic waves. For $R=4$, the same critical value already increases to $q_c=0.276(4)$, whereas for $R=5$, leaving all else the same, the certainty of recurrent epidemic waves never comes to be, with the highest probability being just below $60\%$ at $q=1$, as can be observed in the middle and rightmost panel of Fig.~\ref{quanty} (solid grey line).

Some of these terrible outcomes can be mitigated by imagining even greater control over the newborns that follow nanowar eliminations. For example, one could imagine the newborns being not only cooperators, but also recovered rather than susceptible, i.e., immune at least for the duration of the immunity time, or longer. We have verified that this significantly lowers the probability for recurrent epidemic waves (results not shown), but still does nothing to promote cooperation.

With regards to the latter, the fact that a nanowar fails to promote cooperation can still be considered as a relatively lenient outcome. Research has shown that people are often conditionally cooperative~\cite{fischbacher2001people, frey2004social, grujic2014comparative, thoni2018conditional}, cooperating only if others also cooperate. A model where an eliminated defector would not be replaced by a cooperator but simply left vacant could, thus, easily further erode cooperation in such a case. Many other alterations to the current setup can be envisaged along similar lines. Future research will certainly address some of these possibilities. For the time being, however, we feel that the presented results deter convincingly against considering, let alone implementing, nanowar.

To conclude, we underline, as already emphasized time and again in various works in social physics and computational social science~\cite{castellano_rmp09, jusup_pr22}, that linear thinking often fails when it comes to attempts to better society. As the nanowar thought experiment proves, we must expect the unexpected, such as the fact that killing defectors and replacing them with cooperators can actually impair cooperation rather than promoting it, or that killing off infected agents actually feeds an epidemic rather than halting it. The secret to understanding such counterintuitive results lies in collective behavior and self-organization in complex dynamical systems, which cannot be determined or enforced by top down approaches, but should be promoted in a bottom-up way, by assisting interactions of individuals, which are expected to produce favorable collective effects~\cite{NextCivilization}.

\section*{\normalsize{Discussion}}
\label{discuss}

The aim of our study is to warn of dangers of social engineering and to raise awareness for potential threats by dual uses of nanotechnology. At the moment, the use of nanoparticles seems to be under-regulated. Adding nanoparticles to food, for example, mostly does not have to be declared, and possible societal side effects of using nanotechnology are often not discussed. However, everyone is exposed to nanoparticles via food, water, air, contact, drugs or vaccines in various concentrations, while some of them are toxic~\cite{maher2016magnetite}. Under such conditions, dual uses are conceivable. We also remind of Murphy's law, according to which everything that \textit{can} go wrong \textit{will} go wrong, sooner or later.

To make the case for international regulation, we have chosen a particularly drastic, ``apocalyptic'' kind of scenario, which, if it really took place, would certainly be judged as crime against humanity. Not only do we condemn the implementation of such scenarios. Our results also show that the attempt to improve the state of the world by killing defectors can miserably fail: The existence of a social dilemma situation will cause a never-ending re-occurrence of defectors, where a ``war against misbehaviour'' (here: defection) can even decrease cooperation. This is also expected when the agents are conditionally cooperative, i.e. if they cooperate when having a sufficient number of cooperative neighbors---especially when the population density is reduced. As a consequence, we doubt that a better world will result by targeting individuals and trying to eradicate ``bad behavior'' or even killing defectors.

Instead, for a desirable collective behavior to emerge in a complex dynamical system such as our society, one needs to focus on the interactions between system components rather than the components themselves (the agents) \cite{NextCivilization}. This is a matter of ``mechanism design''. In fact, complexity theory, computational social science, and evolutionary game theory are full of examples illustrating how interactions may be changed in such a way that cooperation is promoted by local interactions.

\section*{\normalsize{Acknowledgments}}
D.H. acknowledges support through the project ``HumanE AI Network'', which has received funding from the European Union's Horizon 2020 research and innovation programme (Grant No. 952026).
M.P. acknowledges support from the Slovenian Research Agency (Grant Nos. P1-0403 and J1-2457).

\clearpage

\clearpage

\clearpage

\end{document}